\documentclass[prl,12pt]{revtex4}

\usepackage[english]{babel}
\usepackage[intlimits]{amsmath}
\usepackage[dvips]{graphicx}

\begin{document}

\title{Molecular Motors Interacting with Their Own Tracks}
\author{Max N. Artyomov}
\affiliation{ Department of Chemistry, Massachusetts Institute of Technology, Cambridge, MA 02139}
\author{Alexander Yu. Morozov and Anatoly B. Kolomeisky}
\affiliation{Department of Chemistry, Rice University, Houston, TX 77005 USA}

\begin{abstract}
Dynamics of molecular motors that move along linear lattices and interact with them via reversible destruction of specific lattice bonds is investigated theoretically by analyzing exactly solvable discrete-state ``burnt-bridge'' models. Molecular motors are viewed as diffusing particles that can asymmetrically break or rebuild  periodically distributed weak links  when passing over them. Our explicit calculations of dynamic properties show that coupling the transport of the unbiased molecular motor  with the bridge-burning mechanism leads to a directed motion  that lowers fluctuations and produces a dynamic transition in the limit of low concentration of weak links. Interaction between the backward biased molecular motor and the bridge-burning mechanism yields a complex dynamic behavior. For the reversible dissociation  the backward motion of the molecular motor is slowed down. There is a change in the direction of the molecular motor's motion for some range of parameters. The molecular motor also experiences non-monotonic fluctuations due to the action of two opposing mechanisms: the reduced activity after the burned sites and locking of large fluctuations. Large spatial fluctuations are observed when two mechanisms are comparable. The properties of the molecular motor are different for the irreversible burning of bridges where the velocity and fluctuations are suppressed for  some concentration range, and the dynamic transition is also observed. Dynamics of the system is discussed in terms of the effective driving forces and transitions between different diffusional regimes.

\end{abstract}

\maketitle

In recent years a significant attention has been devoted to theoretical and experimental studies of molecular motors, both biological and artificial, because of their importance for different non-equilibrium processes in chemistry, physics and biology  \cite{julicher97,reimann02,shirai06,AR07}. A fundamental question related to molecular motors is to understand how externally supplied energy is transformed into mechanical motion on the molecular level. Most of theoretical studies suggest that molecular motors move along spatially asymmetric potentials under the influence of thermal and/or other external forces \cite{julicher97,reimann02}. The majority of biological molecular motors, also known as motor proteins, follow this mechanism by hopping along linear filaments and transforming the energy of hydrolysis of ATP (or related compounds) to produce a mechanical work. An alternative possible mechanism of the molecular motor functioning has also been proposed \cite{porto00,oshanin04,fleishman07}. According to this idea the molecular motor can advance along symmetric potentials via spatially inhomogeneous interactions with underlying linear tracks. Recent experimental observations indicate that a protein  collagenase might utilize this mechanism to power its motion \cite{saffarian04,saffarian06}. The collagenase moves processively along collagen fibrils by using asymmetric collagen proteolysis. The  molecule catalyzes the dissociation of the filament at specific positions, and at the end of the process it is always found on one side of the cleavage site. Between the proteolysis sites the molecule jumps with equal probability in both directions. However, because the collagenase cannot cross backward the already destroyed bond, this leads to an effective biased diffusion in one direction  \cite{saffarian04,saffarian06}. 

Theoretical analysis of the collagenase transport suggested that its dynamics can be well described by so-called ``burnt-bridge'' models (BBM) \cite{saffarian04,saffarian06,mai01,antal05,morozov07,artyomov07,dimer}.  In this approach the molecular motor is viewed as a random walker hopping along the discrete lattice composed of strong and weak bonds. The particle might break the weak links after passing over them, while the strong bonds are not affected.  Although  current theoretical picture provides a reasonable description of collagenase dynamics  \cite{mai01,antal05,morozov07,artyomov07,dimer}, it has a significant conceptual problem by assuming the irreversibility of the burning process. Collagenases are catalytic molecules that accelerate {\it both} forward and backward  proteolysis transitions \cite{AR07}. In addition, only unbiased random walkers have been considered so far. In this paper we analyze the mechanisms of how molecular motors can move along the linear tracks by asymmetrically breaking and rebuilding periodically distributed weak bonds. It is found that coupling between these non-equilibrium processes leads to several unusual phenomena such as dynamic transitions, strong or suppressed fluctuations, and the reversal of the motion direction.  

In our model we consider a molecular motor molecule as a random walker that translocates along the one-dimensional lattice, as shown in Fig 1. The lattice consists of strong and weak links (bridges). The weak bonds are periodically distributed, i.e., the concentration of bridges is $c=1/N$. The molecular motor always moves with the rate $u$ ($w$) to the right (left) while passing over the strong links. However, the dynamic rules for crossing the weak links are different. When the particle crosses the weak link in the forward direction  (from left to right) it does it with the rate $u$, and the bridge can be burned with the probability  $p_{1}$. The rate of the backward transition via the weak bond is $w$. The particle after the already burnt bridge (for example, at the position 0 in Fig. 1) might attempt to move backward.  The bridge can be recovered with the probability $p_{2}$,  and the particle moves one site to the left with the rate $w$. However, with the probability $(1-p_{2})$ the attempt might fail, and then the particle stays at the same site. Note that in the case of $p_{2}=1$ there is no coupling of the molecular motor motion with the bridge-burning  processes. 

Explicit calculations of dynamic properties of molecular motors in reversible BBM can be performed using methods developed earlier \cite{morozov07,artyomov07,dimer}. However, to highlight the physics of interaction between  molecular motor's transport and dissociation of weak links  we consider a simpler case of $p_{1}=1$ and arbitrary $p_{2}$, which corresponds to deterministic burning and stochastic recovery of bridges. Then the motion of the molecular motor can be viewed as hopping on the periodic lattice (with the period $N$). All forward rates  are equal to $u$, while the backward rates are $w$ with the exception of the crossing the burned bridge that has the rate $w p_{2}$: see Fig. 1. The dynamic properties of a single particle on periodic lattices are known exactly \cite{derrida83}, and these results can be used to compute the properties of the molecular motor in reversible BBM. The mean velocity for $p_{1}=1$ is given by
\begin{equation}\label{eq_v}
V=\frac{w(p_{2} - \beta^{1/c})(1- \beta)^{2}}{(\beta^{1/c}-p_{2})(1-\beta) -c (1-\beta^{1/c}) (1-p_{2})},
\end{equation}
where $\beta=(u/w)$. Similar calculations can be done for the dispersion $D$. Specifically, for $u=w=1$ one can obtain
\begin{equation}\label{eq_D1}
D=\frac{2}{3}\frac{(1-p_{2})^{2}(1-p_{2}+2/c+2p_{2}/c)+(1/c)^{2}(1+4p_{2}+p_{2}^{2})(2-2p_{2}+1/c +p_{2}/c)}{(1-p_{2}+1/c+p_{2}/c)^{3}},
\end{equation}
while for the general rates $u$ and $w$ and for $p_{2}=0$ it can be shown that
\begin{equation}\label{eq_D2}
D=\frac{u (1-\beta)^{2} \beta^{1/c} \left[(c/\beta)(1-\beta^{1/c})(1+\beta^{1/c}+4\beta^{1+1/c})-(1-\beta)(\beta^{-1+2/c}+\beta^{2/c}+4\beta^{-1+1/c}) \right]}{2\left[c(1-\beta^{1/c})-\beta^{1/c}(1-\beta)\right]^{3}}.
\end{equation}

To investigate the effect of the reversible burning of bridges on dynamics of molecular motors first we consider the unbiased particle hopping with the rates $u=w$. The results for the dynamic properties are presented in Fig. 2. After burning the bridge the molecular motor  has a non-zero probability of not moving to the left, while it can always freely hop to the right. This interaction between the unbiased molecular motor and the linear track produces an effective force that breaks the symmetry and drives the particle in the forward direction. As expected, increasing  the concentration of weak bonds and/or lowering the recovery probability $p_{2}$  accelerates the particle motion: see Fig. 2a. Simultaneously, it lowers  particle fluctuations because the mobility of the molecular motor is decreased at the sites  after  the dissociated bond. However, an usual behavior is observed in the dispersion in the limit of very low concentration of bridges. For the lattice without weak links one expects $D(c=0)=(u+w)/2=u$, but the dispersion of the particle in reversible BBM goes to another  limit. For example, for $u=w=1$  from Eq. (\ref{eq_D1}) it can be shown that
\begin{equation}
D(c\rightarrow 0)= \frac{2(1+4p_{2}+p_{2}^{2})}{3(1+p_{2})^{2}},
\end{equation}
which is not equal to $D(c=0)=1$ for $0 \le p_{2} <1$. There is always a gap in the dispersion in the limit of $c \rightarrow 0$. This jump in the dispersion is an indication of a dynamic transition between two  regimes: for $c=0$ the molecular motor experiences the unbiased diffusion, while for any non-zero concentration of weak bonds the system is in the biased diffusion regime. Symmetry breaking is an important part of this dynamic transition.

More interesting dynamics is observed when the molecular motor that diffuses to the left is put on the lattice with weak links that generate an effective force in the opposite direction. The dynamic properties for the system are given in Fig. 3. For any non-zero probability of  recovery  the mean velocity of the molecular motor is increasing as a function of the concentration of bridges, i.e., the particle moves slower in the backward  direction (see Fig. 3a). For some values of $p_{2}$ the driving force by reversible burning of bridges becomes so large that it changes the direction of the particle's motion. The critical concentration $c^{*}$ at which it takes place can be found from Eq. (\ref{eq_v}), $c^{*}=\ln{\beta}/\ln{p_{2}} $. It also indicates that the direction reversal can be observed only when $p_{2} \le \beta=u/w$. At the critical concentration the effective force of the asymmetric burning mechanism becomes equal to the backward force of the original molecular motor's motion \cite{AR07}. When the bridge burning becomes irreversible ($p_{2}=0$), the mean velocity behaves differently. The particle always moves forward even  for the smallest concentration of weak bonds. It can be derived from Eq. (\ref{eq_v}) that
\begin{equation}
V(p_{2}=0)=\frac{w\beta^{1/c}(1- \beta)^{2}}{c (1-\beta^{1/c}) -\beta^{1/c}(1-\beta)}>0
\end{equation}
for all non-zero concentrations, which differs from $V(c=0)=u-w<0$. This observation can be explained by looking at the details of the particle's motion. The molecular motor mostly moves backward, but sometimes it diffuses in the forward direction. Crossing the weak bond breaks it and the particle becomes locked in a new position because of irreversibility, creating the effective motion to the right. In the limit of very low concentrations the velocity is decaying exponentially to zero, $V \simeq w \beta^{1/c} (1-\beta)^{2}/c \rightarrow 0$. This result suggests that the bridge-burning mechanism starts to work efficiently only for concentrations larger than $\ln{(1/\beta)}=\ln{(w/u)}$ after overcoming the driving force of the backward biased diffusion.

The behavior of the dispersion of the backward diffusing molecular motor on the lattice with weak bonds is more complex, as shown in Fig. 3b. Although the particle mostly jumps to the left, sometimes fluctuations in the positive direction bring it to the right close to the next weak bond.  After burning the bridge the motion of the particle to the left is significantly reduced, and this large positive fluctuation is locked, leading to increase in the dispersion as a function of $c$. However, the molecular motor sitting after the burned bridge experiences lower mobility that reduces its dispersion. The interplay of these two mechanisms explains the behavior of the dispersion. For low recovery probabilities $p_{2}$ and low concentrations of bridges the locking mechanism dominates, while for large $p_{2}$ and $c$ the reduced mobility is the main factor affecting fluctuations. Surprisingly, we found that when the contributions of both mechanisms in the dynamics of the molecular motor are similar the particle might experience very large fluctuations as $p_{2} \rightarrow 0$. For the irreversible burning of bridges  ($p_{2}=0$) the locking mechanism dominates at all concentrations. In the limit of  $c \rightarrow 0$ one can show from Eq. (\ref{eq_D2}) that $D \simeq w \beta^{1/c} (1-\beta)^{2}/(2c^{2}) \rightarrow 0$, i.e., the fluctuations are suppressed for concentrations less than $\ln{(1/\beta)}=\ln{(w/u)}$ until the effective force of the burning mechanism overcomes the backward driving force of the molecular motor. Again, for the irreversible burning and $c=0$ there is a dynamic transition. In this case, both $V$ and $D$ have gaps at $c=0$: see Fig. 3. This transition separates the backward biased ($c=0$) from the forward biased motion (for $c \ne 0$). It is interesting to note that for the finite probability of recovery, $p_{2}>0$, there are no indications of the dynamic transition. At the critical concentration $c^{*}$ the dispersion is a smooth function without any singularity.

In conclusion, we have investigated the transport of molecular motors that interact with their own linear tracks using discrete lattice models that allowed us to calculate explicitly the dynamic properties of the system. The coupling between non-equilibrium processes of the molecular motor motion and dissociation of weak links on the lattice leads to a complex dynamic behavior with dynamic transitions, inversion of the direction of the motion, and increasing or suppressing fluctuations. The asymmetric burning of bridges creates an effective force that stimulates the motion of the molecular motor in the forward direction. For the unbiased molecular  motor  the interaction with weak links increases the velocity and decreases the dispersion as a function of the concentration of bridges. Similar trends are observed for increasing the non-equilibrium character of the weak bond dissociation by lowering the recovery probability. The dynamic transition between biased and unbiased diffusional regimes has been found in the limit of low concentration of bridges, as indicated by jumps in the dispersions. The coupling of the backward moving molecular motor with the forward-directed burning mechanism lowers the backward velocity with increasing the concentration of bridges, and for some recovery probabilities the direction of the motion even can  be reversed. It has been shown that the behavior of fluctuations in the system can be explained by two dynamic mechanisms: 1) lowering the particle's mobility at the sites after the burned bridge due to lower probability of recovery that decreases the fluctuations; and 2) locking large fluctuations that increases the dispersion. The interplay between these mechanisms leads to a non-monotonic behavior in the dispersion with very large spatial fluctuations when contributions from both mechanisms are similar.  The dynamics of the system is different for the irreversible burning of weak bonds. In this case the dynamics is governed by locking of fluctuations mechanism that also produces jumps in the velocity and in the dispersion in the limit of very low concentration of bridges. For large range of concentrations the velocity and fluctuations are suppressed until the effective force of the burning mechanism overcomes the backward diffusion force of the molecular motor. Thus the dynamics of molecular motors strongly depends on the degree of irreversibility of the coupled bridge-burning mechanism. It is important to note that the dynamics of the single molecular motors has been analyzed in this paper. However, it will be important to investigate the effect of the coupling of many molecular motors with  asymmetric dissociation of links in order to better understand the fundamental nature of these non-equilibrium processes.

The support  from the Welch Foundation (under Grant No. C-1559), and from the US National Science Foundation (grants CHE-0237105 and NIRT ECCS-0708765) is gratefully acknowledged.

\newpage

\noindent {\bf Figure Captions:} \\\\

\noindent Fig. 1. A schematic picture of the reversible burnt-bridge models for the transport of molecular motors. Thick solid lines represent strong bonds, while thin solid lines are for weak bonds. Dotted lines correspond to already destroyed links. Weak bonds are periodically distributed every $N$ sites. 

\vspace{5mm}

\noindent Fig. 2. Dynamic properties of the unbiased molecular motor with $u=w=1$: a) The mean velocity as a function of the concentration of bridges for different recovery probabilities; b) The dispersion  as a function of the concentration of bridges for different recovery probabilities.

\vspace{5mm}

\noindent Fig. 3. Dynamic properties of the backward biased molecular motor with $u=0.2$ and $w=0.8$: a) The mean velocity as a function of the concentration of bridges for different recovery probabilities; b) The dispersion  as a function of the concentration of bridges for different recovery probabilities.

\newpage

\noindent \\\\\\

\begin{figure}[ht]
\unitlength 1in
\resizebox{3.375in}{1in}{\includegraphics{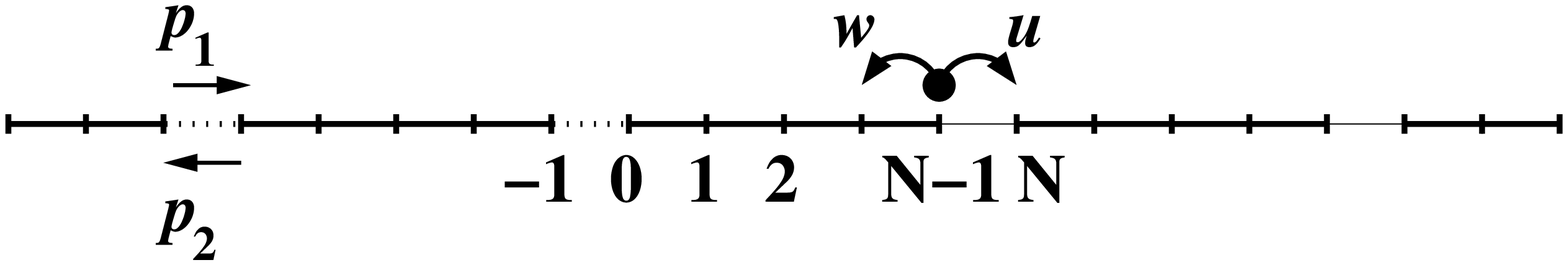}}
\vskip 0.3in
\caption{{\bf  Artyomov, Morozov and Kolomeisky, Physical Review Letters.}}
\end{figure}

\newpage

\noindent \\\\\\

\begin{figure}[ht]
\unitlength 1in
[a] \ \resizebox{3.0in}{2.2in}{\includegraphics{Fig2a.eps}}\\
\vskip 1in 
[b] \ \resizebox{3.0in}{2.2in}{\includegraphics{Fig2b.eps}}
\vskip 1in
\caption{{\bf  Artyomov, Morozov and  Kolomeisky, Physical Review Letters.}}
\end{figure}

\newpage

\noindent \\\\\\

\begin{figure}[ht]
\unitlength 1in
[a] \ \resizebox{3.0in}{2.20in}{\includegraphics{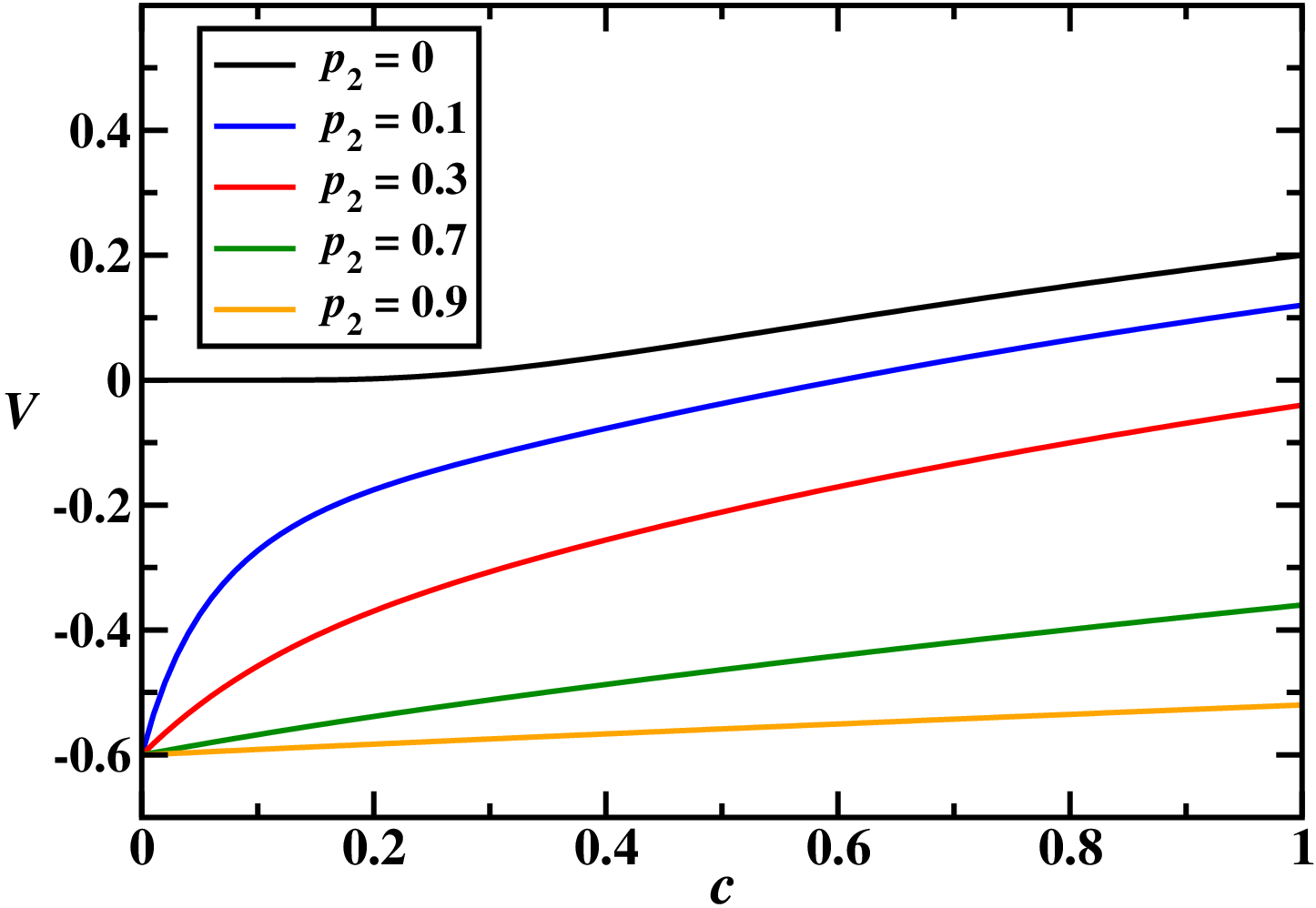}}\\
\vskip 1in 
[b] \ \resizebox{3.0in}{2.25in}{\includegraphics{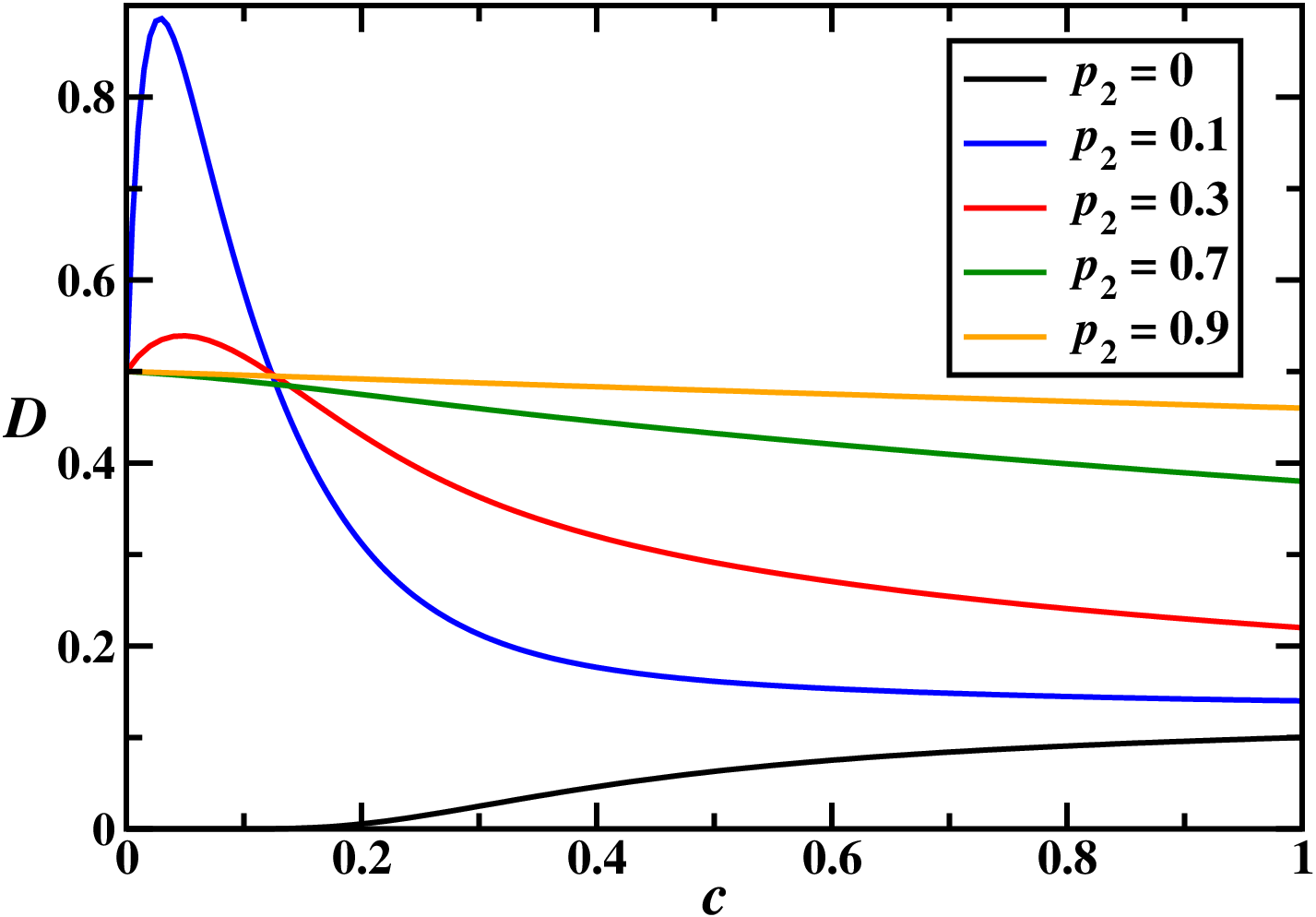}}
\vskip 1in
\caption{{\bf  Artyomov, Morozov and  Kolomeisky, Physical Review Letters.}}
\end{figure}

\end{document}